\pdfoutput=1
\documentclass[12pt]{iopart} 
\expandafter\let\csname equation*\endcsname\relax
\expandafter\let\csname endequation*\endcsname\relax
\usepackage[version=4]{mhchem}
\usepackage{graphicx}
\usepackage{xcolor}

\begin{document}
\title[Theory of tunnel magnetoresistance in magnetic tunnel junctions with 
hexagonal]{Theory of tunnel magnetoresistance in magnetic tunnel junctions with 
hexagonal boron nitride barriers: mechanism and application to ferromagnetic alloy electrodes
}
\author{I Kurniawan$^1$, K Masuda$^1$, and Y Miura$^{1,2}$}
\address{$^1$ Research Center for Magnetic and Spintronic Materials, National Institute for Materials Science (NIMS), Tsukuba 305-0047, Japan}
\address{$^2$ Faculty of Electrical Engineering and Electronics, Kyoto Institute of Technology, Matsugasaki, Sakyo-ku, Kyoto, 606-8585, Japan}
\ead{kurniawan.ivan@nims.go.jp}

\begin{abstract}
Hexagonal boron nitride (\textit{h}-BN), with its strong in-plane bonding and good lattice match to hcp and fcc metals, offers a promising alternative barrier material for magnetic tunnel junctions (MTJs). Here, we investigate spin-dependent transport in \ce{hcp-Co_{1–\it{x}}Ni_{x}}$/$\textit{h}-BN$/$\ce{hcp-Co_{1–\it{x}}Ni_{x}}(0001) MTJs with physisorption-type interfaces using first-principles calculations. We find that a high TMR ratio arises from the resonant tunneling of the down-spin surface states of the \ce{hcp-Co_{1–\it{x}}Ni_{x}}, having a $\Delta_1$-like symmetry around the $\Gamma$ point. Ni doping tunes the Fermi level and enhances this effect by reducing the overlap between up-spin and down-spin conductance channels in momentum space under the parallel configuration, thereby suppressing antiparallel conductance and increasing the TMR ratio. This mechanism is analogous to Brillouin zone spin filtering and is sensitive to the interfacial distance but not specific to \textit{h}-BN barriers; similar behavior may emerge in MTJs with other two-dimensional insulators or semiconductors. These findings provide insight into surface-state-assisted tunneling mechanisms and offer guidance for the interface engineering of next-generation spintronic devices.
\end{abstract}
\maketitle

\section{Introduction}

The tunneling magnetoresistance (TMR) effect, characterized by a resistance change in magnetic tunnel junctions (MTJs), has been central to spintronics since its first low-temperature observation 50 years ago by Jullière~\cite{julliere1975}. An MTJ typically consists of a trilayer structure: ferromagnetic (FM) electrode$/$non-magnetic insulator (NM)$/$FM electrode. The relative magnetization direction of the FM layers modulates the resistance through spin-dependent tunneling, which is essential for applications such as hard disk drives (HDDs) and magnetoresistive random-access memories (MRAMs)~\cite{hirohata2020}. Over the past decades, efforts to improve the TMR ratio have focused on optimizing the spin polarization of FM electrodes and the symmetry selectivity of NM barrier materials, while maintaining lattice matching at FM$/$NM interfaces. In parallel, theoretical studies have significantly advanced our understanding of spin-dependent tunneling and have guided the design of high-performance MTJs.

Early magnetic tunnel junctions (MTJs) employed amorphous \ce{Al_{2}O_{3}} barriers, leading to the first room-temperature TMR observations by Miyazaki \textit{et al.}~\cite{miyazaki1995} and Moodera \textit{et al.}~\cite{moodera1995} in 1995. In such amorphous barriers, the tunneling conductance can be approximately described by the density of states of the ferromagnetic electrodes, as captured by Jullière’s model~\cite{julliere1975}. However, theoretical studies based on first-principles calculations by Butler \textit{et al.}~\cite{butler2001} and Mathon \textit{et al.}~\cite{mathon2001} revealed that tunneling conductance in crystalline Fe$/$MgO$/$Fe(001) MTJs exhibits more complex behavior. In particular, the conductance is highly symmetry-dependent, with different wavefunction symmetries decaying at different rates within the barrier. The dominant contribution arises from $\Delta_1$ states, characterized by $s$, $p_z$, and $d_{3z^2 - r^2}$ orbital characters, which correspond to the slowest-decaying evanescent states in the complex band structure of MgO. These $\Delta_1$ states are fully spin-polarized in bcc Fe, resulting in a high TMR ratio, as predicted by Butler and Mathon~\cite{butler2001,mathon2001}. This theoretical prediction was later realized experimentally by Yuasa \textit{et al.}~\cite{yuasa2004} and Parkin \textit{et al.}~\cite{parkin2004} in 2004.

Afterwards, many efforts were devoted to improving the MTJ structure or discovering better materials to obtain higher TMR ratios. From the perspective of materials design, some Heusler alloys are predicted to be half-metallic ferromagnets such as \ce{Co2MnSi}~\cite{galanakis2002}, meaning they are conductive in only one spin channel. They are expected to exhibit huge TMR ratios through coherent tunneling~\cite{miura2008}. However, such tunneling requires lattice-matched junctions, which is a challenge for MgO-based MTJs due to the significant lattice mismatch even with conventional ferromagnets such as Fe (3.5\%),  CoFe (3.7\%), and Heusler alloys \ce{Co2MnSi} (5.1\%)~\cite{sukegawa2010}. This mismatch introduces interfacial defects that degrade spin-dependent tunneling and suppress TMR ratio. To address this issue, alternative NM layers such as spinel oxides (\ce{MgAl2O4}~\cite{sukegawa2010}, \ce{MgGa2O4}~\cite{sukegawa2017}) and interface engineering techniques, including ultrathin layer insertion~\cite{scheike2021}, have been explored. Most recently, CoFe$/$MgO$/$CoFe(001) MTJs reached a record 631\% TMR ratio at room temperature in 2023~\cite{scheike2023}. However, these advances also highlight a fundamental limitation of MgO-based systems, which is their sensitivity to lattice matching and interfacial quality. These factors constrain the maximum achievable TMR in practical devices.

In 2004, the isolation of monolayer graphene spurred the exploration of other two-dimensional (2D) materials for spintronic applications, including hexagonal boron nitride (\textit{h}-BN)~\cite{novoselov2004}. Yazyev and Pasquarello later proposed graphene and \textit{h}-BN as atomically thin spacers for magnetoresistive junctions, and calculated the TMR ratio and conductance using first-principles methods~\cite{yazyev2009}. However, monolayer spacers—particularly graphene, due to its semimetallic nature, and \textit{h}-BN, due to its extreme thinness—tend to exhibit low TMR ratios and low resistance, partly because of in-gap states associated with ultrathin barriers. To overcome these limitations, insulating \textit{h}-BN with increased thickness has emerged as a promising candidate, offering chemical inertness and a small lattice mismatch (less than 1\%) with ferromagnetic (111) or (0001) surfaces. Experimental advances followed, including the work by Dankert \textit{et al.}, who reported a TMR ratio of 0.5\% using transferred \textit{h}-BN on permalloy and Co electrodes~\cite{dankert2015}. Subsequently, Piquemal-Banci \textit{et al.} demonstrated a much higher TMR ratio of 50\% at low temperature using directly grown \textit{h}-BN with Co and Fe electrodes~\cite{banci2018}, although details of the growth process were not fully disclosed. More recently, Emoto \textit{et al.} reported large-area growth of few-layer \textit{h}-BN, observing modest TMR ratios of around 10\% for three-monolayer devices, higher than those of two-monolayer counterparts~\cite{emoto2024}. \textcolor{black}{Beyond its TMR properties, hcp-Co/\textit{h}-BN heterostructures has also been predicted to exhibit substantial perpendicular magnetic anisotropy~\cite{hastuti2025}, which is beneficial for future spintronic sensor applications.}

From a theoretical perspective, Faleev \textit{et al.} predicted a significantly high TMR ratio below the Fermi level in hcp-Co$/$\textit{h}-BN$/$hcp-Co(0001), attributed to a Brillouin zone spin-filtering mechanism~\cite{faleev2015}. This mechanism relies on the presence of high-transmission regions that coincide with available states in only one spin channel within the two-dimensional Brillouin zone corresponding to the in-plane periodicity of MTJs. In essence, it is analogous to the coherent tunneling mechanism in MgO-based MTJs~\cite{butler2001,mathon2001}, but with a key distinction: in \textit{h}-BN-based MTJs, metallic transmission occurs exclusively in one spin channel, particularly near the K point close to the valence band maximum of \textit{h}-BN. Consequently, the TMR ratio is expected to exhibit an exponential increase with respect to the barrier thickness~\cite{faleev2015}, in contrast to the linear or quadratic scaling observed in MgO-based MTJs. However, realizing this behavior requires hole doping of \textit{h}-BN to align its valence band maximum with the Fermi level of the cobalt electrode, which remains experimentally challenging. On the other hand, recent first-principles calculations show that the interfacial distance between the \textit{h}-BN barrier and the hcp-Co electrode significantly affects the TMR ratio~\cite{lu2021acs, lu2021apr, robertson2023}.

Despite these insights, the tunneling mechanism of \textit{h}-BN based MTJs near the Fermi level remains unclear, particularly under different interfacial distances. In addition, systematic investigations into how key parameters, such as energy dependence and orbital symmetry, contribute to TMR ratio enhancement have not been thoroughly explored, even for simple prototype junctions like hcp-Co$/$\textit{h}-BN$/$hcp-Co(0001). Therefore, in this work, we aim to clarify the detailed mechanism of tunneling transport in hcp-Co$_{1-x}$Ni$_x$$/$\textit{h}-BN$/$hcp-Co$_{1-x}$Ni$_x$(0001) junctions using first-principles calculations. Specifically, we investigate (i) the effect of orbital symmetry on the tunneling mechanism, (ii) the shift of the Fermi level induced by Ni substitution, and (iii) the influence of interfacial distance on spin-dependent tunneling. Our findings provide a more comprehensive understanding of spin-dependent tunneling in \textit{h}-BN-based MTJs and offer practical guidance for designing material interfaces and electrode compositions to enhance TMR performance in future 2D-material-based spintronic devices.

\section{Computational details}
\begin{figure}
\centering
\includegraphics[width=\columnwidth]{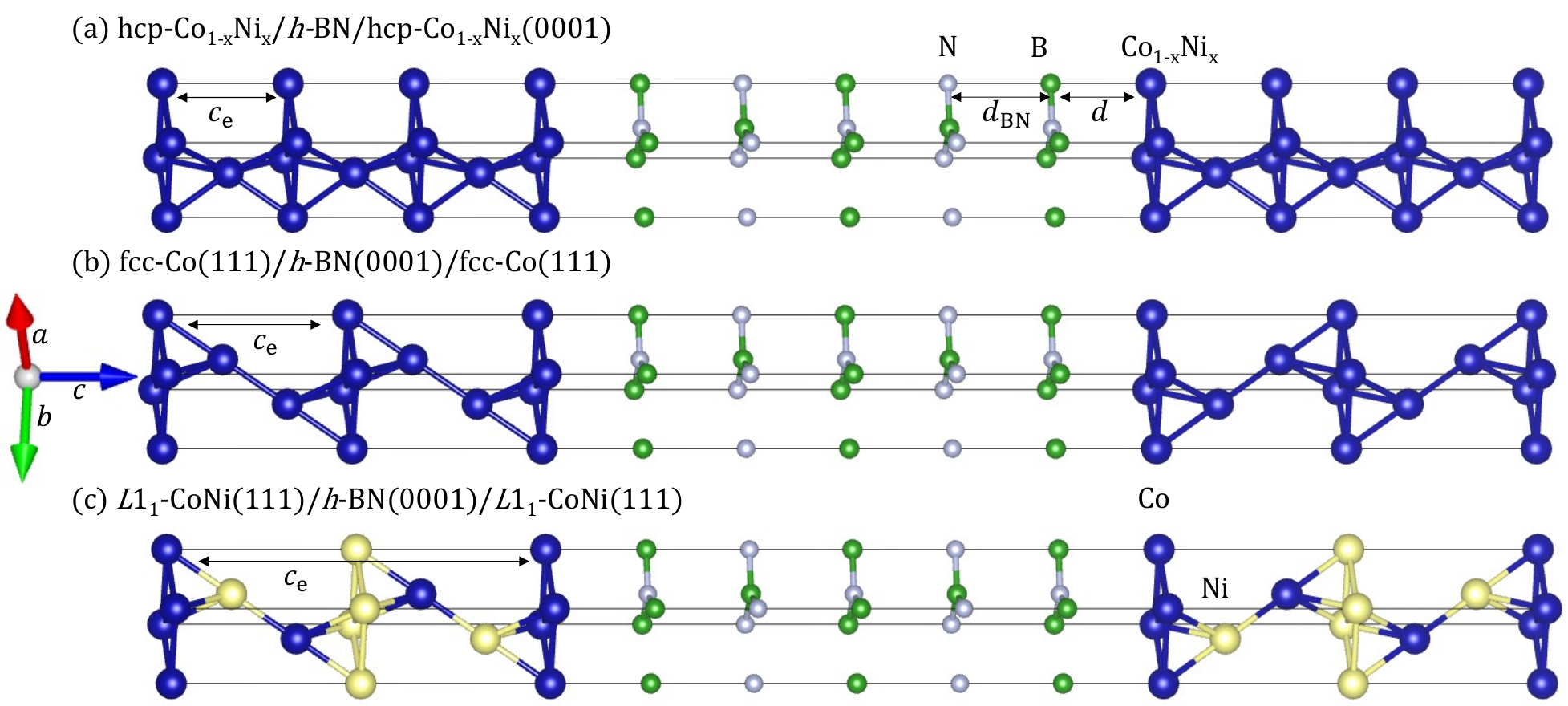}
\caption{\label{fig1} 
Supercells of \textit{X}$/$\textit{h}-BN$/$\textit{X} junctions, where \textit{X} = (a) hcp-\ce{Co_{1-x}Ni_{x}} (top), (b) fcc-Co (middle), and (c) $L1_1$-CoNi (bottom). \textcolor{black}{The $c_\mathrm{e}$ denotes the lattice constant of the ferromagnetic electrode $X$ along the $c$ axis. The $d_\mathrm{BN}$ represents the interlayer spacing of the \textit{h}-BN monolayers, and $d$ is the interfacial distance between \textit{h}-BN and the $X$ electrode.}
}
\end{figure}
We constructed model junctions of hcp-Co$/$\textit{h}-BN$/$hcp-Co(0001) [\Fref{fig1}(a)], a prototype system previously predicted to exhibit a high TMR ratio~\cite{faleev2015}, where the (0001) directions of both the \textit{h}-BN barrier and hcp-Co electrodes are aligned with the stacking direction of the MTJ. The in-plane lattice constants were fixed at $a = b = 2.5$~\AA, based on experimental values for bulk \textit{h}-BN~\cite{lynch1966}, while the interlayer spacing of \textit{h}-BN was optimized using the Tkatchenko–Scheffler (TS) van der Waals correction scheme~\cite{tkatchenko2009}, yielding a layer separation of $d_\mathrm{BN} = 3.33$~\AA. Two representative interface configurations were considered: B-top and N-top, where boron or nitrogen atoms are positioned atop surface metal sites of the Co electrodes~\cite{lu2021acs}. Previous studies have identified that two distinct adsorption regimes of FM on \textit{h}-BN: chemisorption and physisorption~\cite{nagashima1995,rokuta1997}. Chemisorption is associated with shorter interfacial distances and  stronger binding, whereas physisorption features larger interfacial separations and weaker interaction with the substrate. It has also been reported that the B-top configuration is stable for physisorption interfacial distances of 3.0–3.5~\AA\ and can sustain high TMR ratio for barrier thicknesses of 5 monolayers (ML), whereas the N-top configuration is more stable around 2.0–2.5~\AA\ (chemisorption regime) but generally leads to lower TMR ratios~\cite{lu2021acs,lu2021apr,robertson2023}. To represent both regimes, we adopted interfacial distances of $d = 3.25$~\AA\ (physisorption) and $d = 2.15$~\AA\ (chemisorption), with a fixed 5~ML thick \textit{h}-BN barrier in the B-top configuration. In addition to the hcp-Co junction, we also constructed analogous junctions using fcc-Co(111) and $L1_1$-CoNi(111) electrodes [\Fref{fig1}(b–c)] to assess the influence of electrode crystal structure.

The TMR ratio was evaluated using the ballistic transport framework, where conductance is calculated via the Landauer formalism combined with density functional theory (DFT)~\cite{choi1999}. These calculations were performed using the {\scriptsize PWCOND} module~\cite{smogunov2004} implemented in the {\scriptsize QUANTUM ESPRESSO} package~\cite{giannozzi2017}. 
To construct the transport model, we embedded a scattering region comprising the junction \textit{X}$/$\textit{h}-BN$/$\textit{X} between semi-infinite electrodes of material \textit{X}, where \textit{X} represents hcp-\ce{Co_{1-x}Ni_{x}}, fcc-Co, or $L1_1$-CoNi. The electronic structure and scattering potential were obtained through self-consistent DFT calculations using the generalized gradient approximation (GGA)~\cite{perdew1996} and norm-conserving pseudopotentials~\cite{hamann2013}. \textcolor{black}{The choice of GGA is justified by its ability to reproduce the correct qualitative alignment between the Fermi level and the \textit{h}-BN band gap, consistent with the quasiparticle self-consistent GW (QSGW) results of Faleev \textit{et al.}~\cite{faleev2015}, while remaining computationally feasible for systematic alloying studies.} To simulate the atomic disorder effect between Co and Ni in hcp-\ce{Co_{1-x}Ni_x} electrodes, we employed the virtual crystal approximation (VCA)~\cite{bellaiche2000} for the norm-conserving pseudopotentials generated from the PseudoDojo library~\cite{setten2018}, allowing systematic Fermi level tuning and analysis of its influence on spin-dependent transmission.

The plane-wave energy cutoffs were set to 100~Ry for the wavefunctions and 400~Ry for the charge density. A $10 \times 10 \times 1$ \textbf{k}-point grid was used for Brillouin-zone sampling. Owing to the in-plane periodicity of the junction, the scattering states were decomposed by the in-plane wave vector $\mathbf{k}_\parallel = (k_x, k_y)$, where the $x$-axis aligns with the lattice vector $a$. 
For each spin channel and $\mathbf{k}_\parallel$, the scattering problem was solved by matching wavefunctions and their derivatives at the junction–electrode interfaces \textcolor{black}{following the formalism of Choi and Ihm \cite{choi1999} and Smogunov \textit{et al.} \cite{smogunov2004}}. 
\textcolor{black}{In this approach, the transmission coefficient is evaluated as a function of spin index $\sigma$, in-plane momentum $\mathbf{k}_\parallel$, and electron energy $E$, denoted as $T^{\sigma}(\mathbf{k}_\parallel,E)$.} 
The spin-resolved transmission coefficients \textcolor{black}{$T^{\sigma}(\mathbf{k}_\parallel,E)$} were then used to compute the conductance \textcolor{black}{$G_\sigma(\mathbf{k_\parallel},E)$} via the Landauer formula, \textcolor{black}{$G_\sigma(\mathbf{k_\parallel},E)=(e^2/h)\,\times T^{\sigma}(\mathbf{k}_\parallel,E)$.} 
We evaluated four \textcolor{black}{spin-resolved} conductance components: $G_{\mathrm{P},\uparrow}(\mathbf{k}_\parallel)$, $G_{\mathrm{P},\downarrow}(\mathbf{k}_\parallel)$, $G_{\mathrm{AP},\uparrow}(\mathbf{k}_\parallel)$, and $G_{\mathrm{AP},\downarrow}(\mathbf{k}_\parallel)$, corresponding to the up- ($\uparrow$) and down- ($\downarrow$) spin channels in parallel (P) and antiparallel (AP) magnetic configurations \textcolor{black}{at fixed energy}. 
Here, the up-spin (down-spin) channel is defined as the majority-spin (minority-spin) channel in the left electrode in both the P and AP configurations. 
The total conductance for each configuration was obtained by averaging over the Brillouin zone: e.g., $G_{\mathrm{P},\uparrow} = \sum_{\mathbf{k}_\parallel} G_{\mathrm{P},\uparrow}(\mathbf{k}_\parallel)/N$ with $N = 100 \times 100$. 
Finally, the TMR ratio was evaluated using the optimistic definition: 
$\mathrm{TMR\ (\%)} = 100 \times ({G_\mathrm{P} - G_\mathrm{AP}})/{G_\mathrm{AP}},$
where $G_\mathrm{P}$ and $G_\mathrm{AP}$ are the total conductances for the P and AP states, given by the sum of up- and down-spin channels.

\section{Results and discussion}

\begin{figure}
\includegraphics[width=\columnwidth]{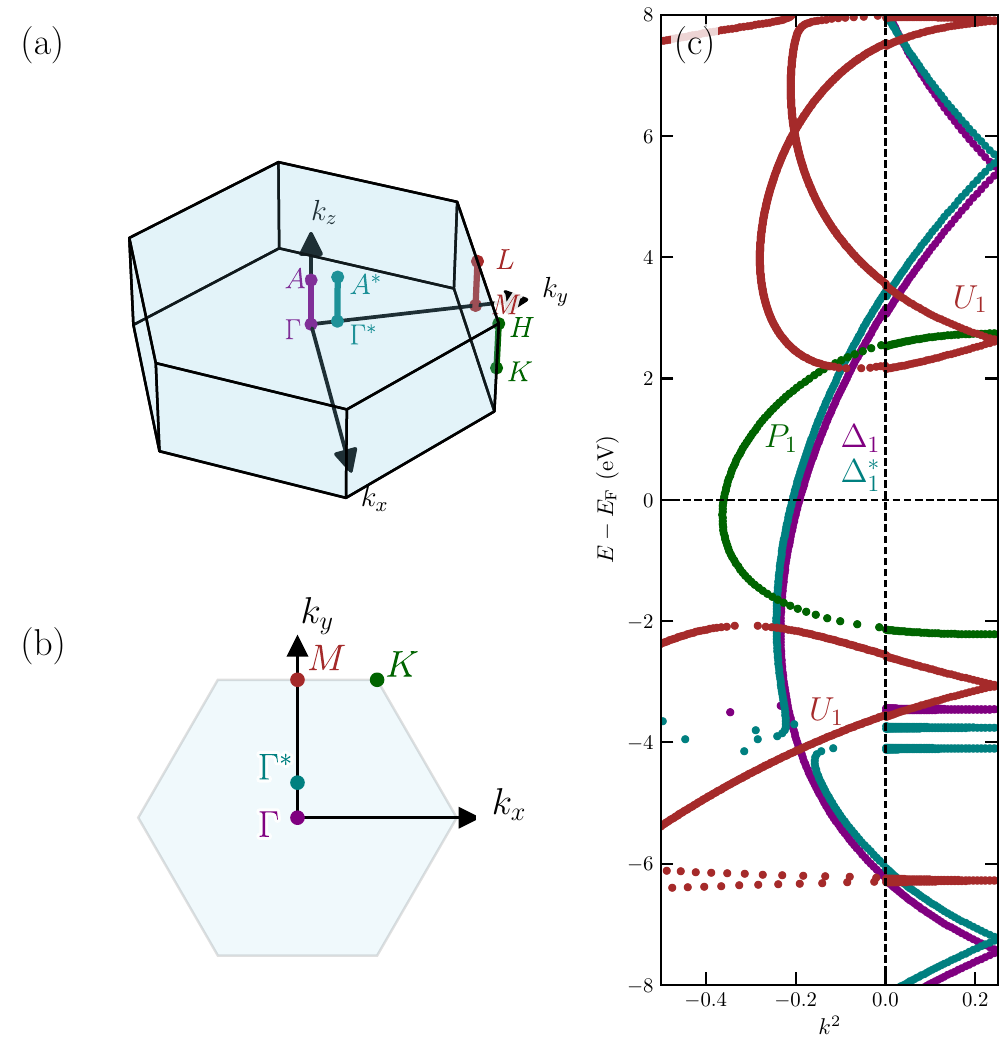}
\caption{\label{fig2} 
(a) Three-dimensional Brillouin zone of the hexagonal lattice.  
(b) Two-dimensional surface Brillouin zone projected onto the (0001) plane.  
(c) Complex band structure of bulk \textit{h}-BN \textcolor{black}{calculated using a unit cell containing two layers, each composed of one B and one N atom}, along several out-of-plane directions: $\Gamma$–A ($\Delta$ line), $\Gamma^*$–A$^*$ ($\Delta^*_1$ line, with $\mathbf{k}_\parallel = (0, 0.1)$), K–H (P line), and L–M (U line). \textcolor{black}{The horizontal axis represents $k^2 = (k_z - k_z^0)^2$ \cite{faleev2015}, where $k_z^0$ is the wave vector corresponding to the lowest conduction-band energy at a given in-plane momentum $\mathbf{k}_\parallel$; $k_z^0 = 0$ for $\mathbf{k}_\parallel = \Gamma, \Gamma^*$, and M, and $k_z^0 = 0.5$ for $\mathbf{k}_\parallel = $K.}
}
\end{figure}

\begin{figure}
\includegraphics[width=\columnwidth]{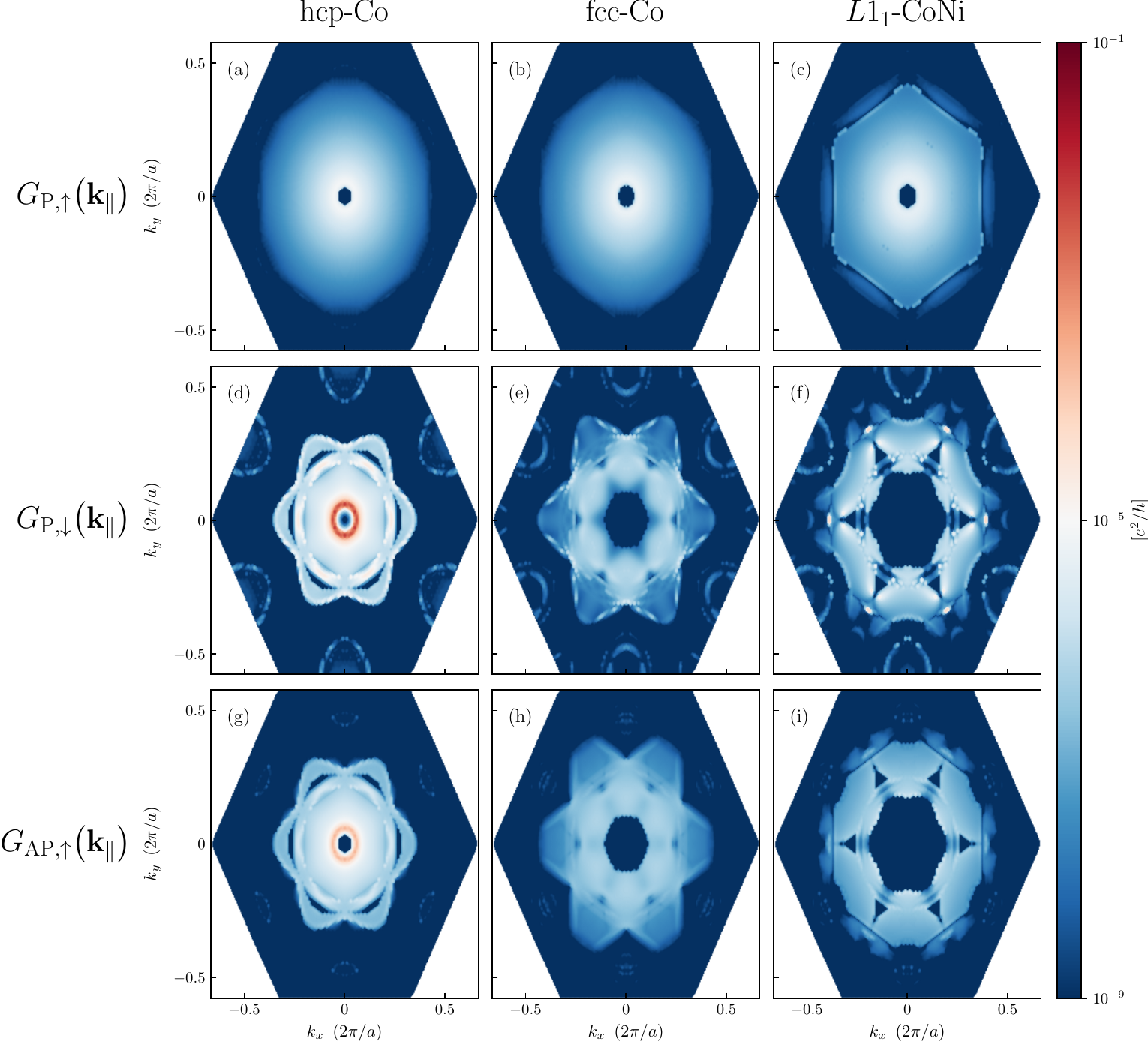}
\caption{\label{fig3} 
$\mathbf{k}_{\parallel}$-resolved conductances at the Fermi level: (a)–(c) up-spin $G_{\mathrm{P},\uparrow}$, (d)–(f) down-spin $G_{\mathrm{P},\downarrow}$ in the parallel configuration, and (g)–(i) up-spin $G_{\mathrm{AP},\uparrow}$ in the antiparallel configuration for \textit{X}$/$\textit{h}-BN$/$\textit{X} junctions with \textit{X} = hcp-Co (a, d, g), fcc-Co (b, e, h), and $L1_1$-CoNi (c, f, i). The unit of conductance is $e^2/h$.
}
\end{figure}

\subsection{Symmetry dependence}
Understanding the TMR effect in \textit{h}-BN–based MTJs remains elusive, partly due to its threefold rotational symmetry, which has been suggested to induce mixing of evanescent states with different orbital symmetries and to exhibit strong $\mathbf{k}_\parallel$-dependent decay rates. Previous studies have indicated that evanescent states with $s$-like character dominate in \textit{h}-BN on high-symmetry lines along $k_z$, such as $\Gamma$–A ($\Delta$ line), K–H (P line), and L–M (U line) [\Fref{fig2}(a)–(b)]. Assuming the Fermi level lies near the center of the \textit{h}-BN band gap, our calculations show that the minimum decay rate still occurs at $\Gamma$ and its vicinity, \textcolor{black}{as inferred from the smallest magnitude of $k^2$, which determines the decay rate via $\exp(-2d\sqrt{-k^2})$ \cite{butler2001}}, even when slightly shifted to $\mathbf{k}_\parallel = (0, 0.1)$ along the $\Gamma^*$–A$^*$ path ($\Delta^*_1$), as shown in \Fref{fig2}(c). Here, the subscript index 1 denotes states that are symmetry-invariant under rotation along the $z$ axis, such as $s$, $p_z$, and $d_{z^2}$ orbitals. Therefore, analyzing the conductance distribution around the $\Gamma$ point remains crucial to elucidating the TMR mechanism in this system. \Fref{fig3} presents the $\mathbf{k}_\parallel$-resolved conductance maps for different electrode materials. A key observation is the down-spin conductance in the parallel configuration for hcp-Co [\Fref{fig3}(d)], which shows a highly conductive region near, but not exactly at, the $\Gamma$ point. This feature contributes significantly to the large down-spin conductance and thus the high TMR ratio in the hcp-Co$/$\textit{h}-BN$/$hcp-Co(0001) junction, as shown in \Tref{table1}, which is consistent with previous calculations~\cite{lu2021acs}. In contrast, no such feature is observed for fcc-Co and $L1_1$-CoNi electrodes [\Fref{fig3}(e)–(f)]. For the up-spin channel in the parallel configuration [\Fref{fig3}(a)–(c)], the conductance does not vary significantly with electrode type. The antiparallel conductances [\Fref{fig3}(g)–(i)] can be qualitatively understood as mixtures of the up- and down-spin channels in the parallel state.

\begin{table}
\caption{\label{table1}Conductance and TMR ratios calculated using supercells \textit{X}$/$\textit{h}-BN$/$\textit{X} junctions with various \textit{X}. The units are in $e^2/h$ and \%, respectively.}
\begin{indented}
\item[]\begin{tabular}{@{}llll}
\br
\textit{X} &hcp-Co&fcc-Co&$L1_1$-CoNi\\
\mr
$G_{\mathrm{P},\uparrow}$&$2.30\times10^{-7}$&$1.85\times10^{-7}$&$1.72\times10^{-7}$\\
$G_{\mathrm{P},\downarrow}$&$1.56\times10^{-5}$&$8.20\times10^{-8}$&$2.51\times10^{-7}$\\
$G_{\mathrm{AP}}$&$2.39\times10^{-6}$&$1.31\times10^{-7}$&$1.26\times10^{-7}$\\
TMR ratio&562&105&238\\
\br
\end{tabular}
\end{indented}
\end{table}

\begin{figure}
\centering
\includegraphics[width=\columnwidth]{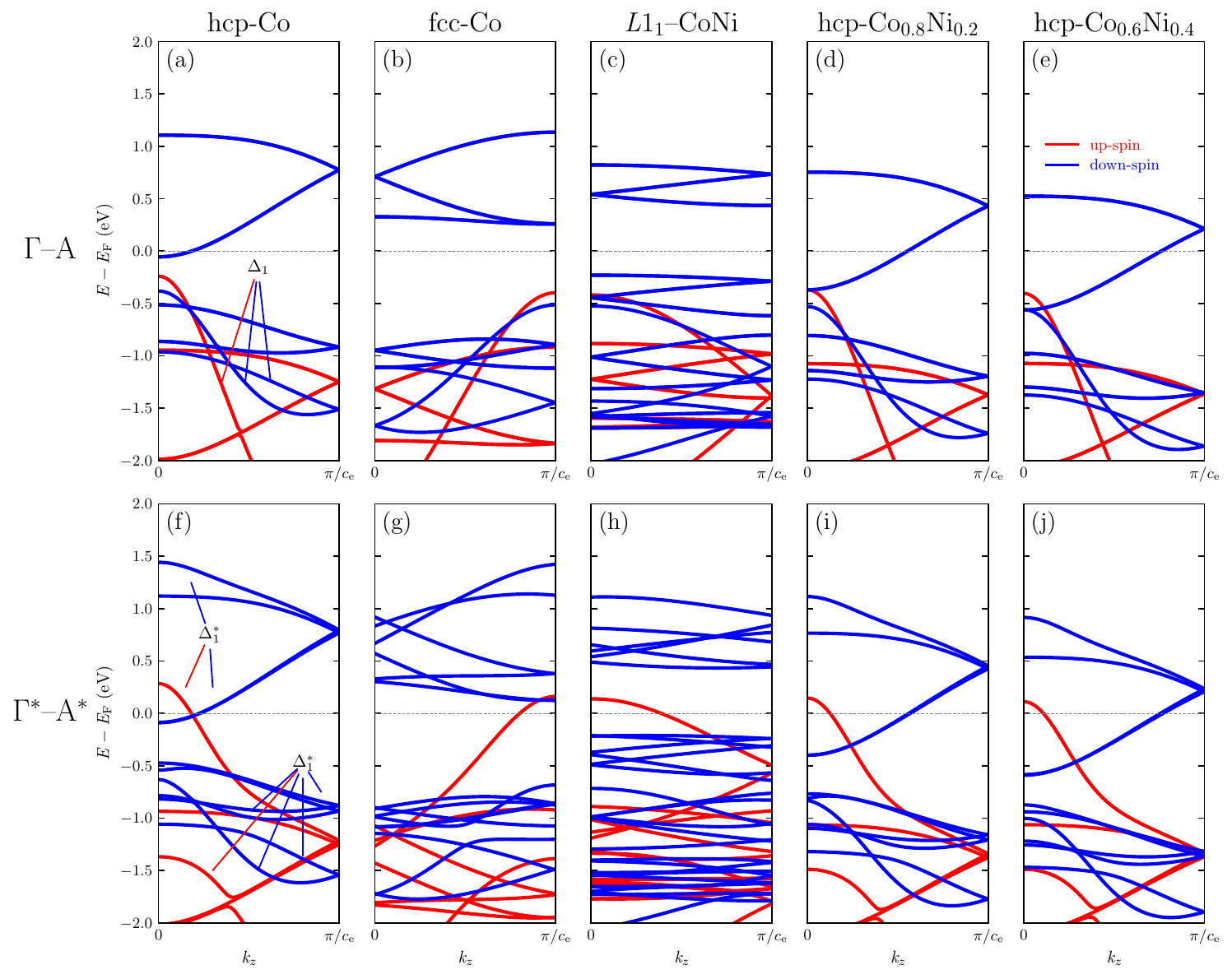}
\caption{\label{fig4} 
Spin-resolved band structures along (a)–(e) the $\Gamma$–A line and (f)–(j) the $\Gamma^*$–A$^*$ line with $\mathbf{k}_\parallel = (0, 0.05)$ for \textit{X} = hcp-Co (a, f), fcc-Co (b, g), $L1_1$-CoNi (c, h), hcp-\ce{Co_{0.8}Ni_{0.2}} (d, i), hcp-\ce{Co_{0.6}Ni_{0.4}} (e, j). Majority- and minority-spin bands are shown in red and blue, respectively. For hcp-Co, bands with $\Delta_1$ and $\Delta^*_1$ symmetry are indicated. \textcolor{black}{The unit of $k_z$, $\pi / c_\mathrm{e}$, is defined using the lattice constant of the ferromagnetic electrode along the $c$ axis, as illustrated in Fig. 1.}
}
\end{figure}

To clarify why the down-spin channel in hcp-Co exhibits a concentrated high-conductance region near the $\Gamma$ point, we examine the band structure of the electrodes in \Fref{fig4}. In the coherent tunneling picture, conductance arises when propagating states in the electrodes couple to evanescent states in the barrier that exhibit the slowest decay, typically those with $\Delta_1$ symmetry along the $\Gamma-\mathrm{A}$ line. In the top panels [\Fref{fig4}(a)–(c)], the spin-resolved band structures along the high-symmetry $\Gamma$–A line are shown. For fcc-Co and $L1_1$-CoNi [\Fref{fig4}(b)-(c)], no bands cross the Fermi level in the minority-spin channel, explaining the absence of conductance near $\Gamma$ [\Fref{fig3}(e)–(f)]. While hcp-Co does have a minority-spin band crossing the Fermi level along $\Gamma$–A [\Fref{fig4}(a)], this state does not have $\Delta_1$ symmetry, consistent with the absence of conductance at $\Gamma$ [\Fref{fig3}(d)]. Interestingly, when shifting to the slightly off-symmetry path $\Gamma^*$–A$^*$ with $\mathbf{k}_\parallel = (0, 0.05)$—corresponding to the location of the observed ring-shaped high-conductance feature in \Fref{fig3}(d)—hcp-Co exhibits a minority-spin band with $\Delta_1$-like symmetry (denoted here as $\Delta^*_1$) crossing the Fermi level [\Fref{fig4}(f)]. This explains the concentrated high conductance in the vicinity of $\Gamma$ point. For fcc-Co and $L1_1$-CoNi, only the majority-spin bands containing $\Delta^*_1$ symmetry cross the Fermi level along this path, consistent with their comparable up-spin conductance and diminished down-spin conductance [\Fref{fig4}(g)-(h)]. 

\subsection{Energy dependence}
In the previous subsection, we identified that the variation in conductance profiles between different electrodes originates from differences in their band structures along lower-symmetry directions. From a practical standpoint, hcp-Co and fcc-Co differ primarily in their stacking sequence, while $L1_1$-CoNi adopts the same stacking as fcc-Co but alternates Co and Ni atoms. This highlights the fabrication challenge of realizing high-performance hcp-Co$/$\textit{h}-BN(0001)-based junctions, since layer-by-layer growth often results in stacking faults or deviations from the ideal hcp structure—conditions under which no $\Delta^*_1$ band crosses the Fermi level, even along off-symmetry paths. To provide a more comprehensive picture of TMR enhancement, we next investigate how tuning the Fermi level, via simulated Ni doping in the electrodes, influences the tunneling conductance. \textcolor{black}{Although metastable hcp-type alloys can be realized under suitable epitaxial growth conditions, as demonstrated for \ce{hcp-Ni_{1–\it{x}}Fe_{\it{x}}} thin films by Huang \textit{et al.}~\cite{huang1998}, our primary objective here is to clarify the effect of Fermi-level tuning within a rigid-band model on the enhancement of the TMR ratio, rather than to propose Ni-doped hcp-Co as a practical electrode material.}

\begin{figure}
\centering
\includegraphics[width=0.5\columnwidth]{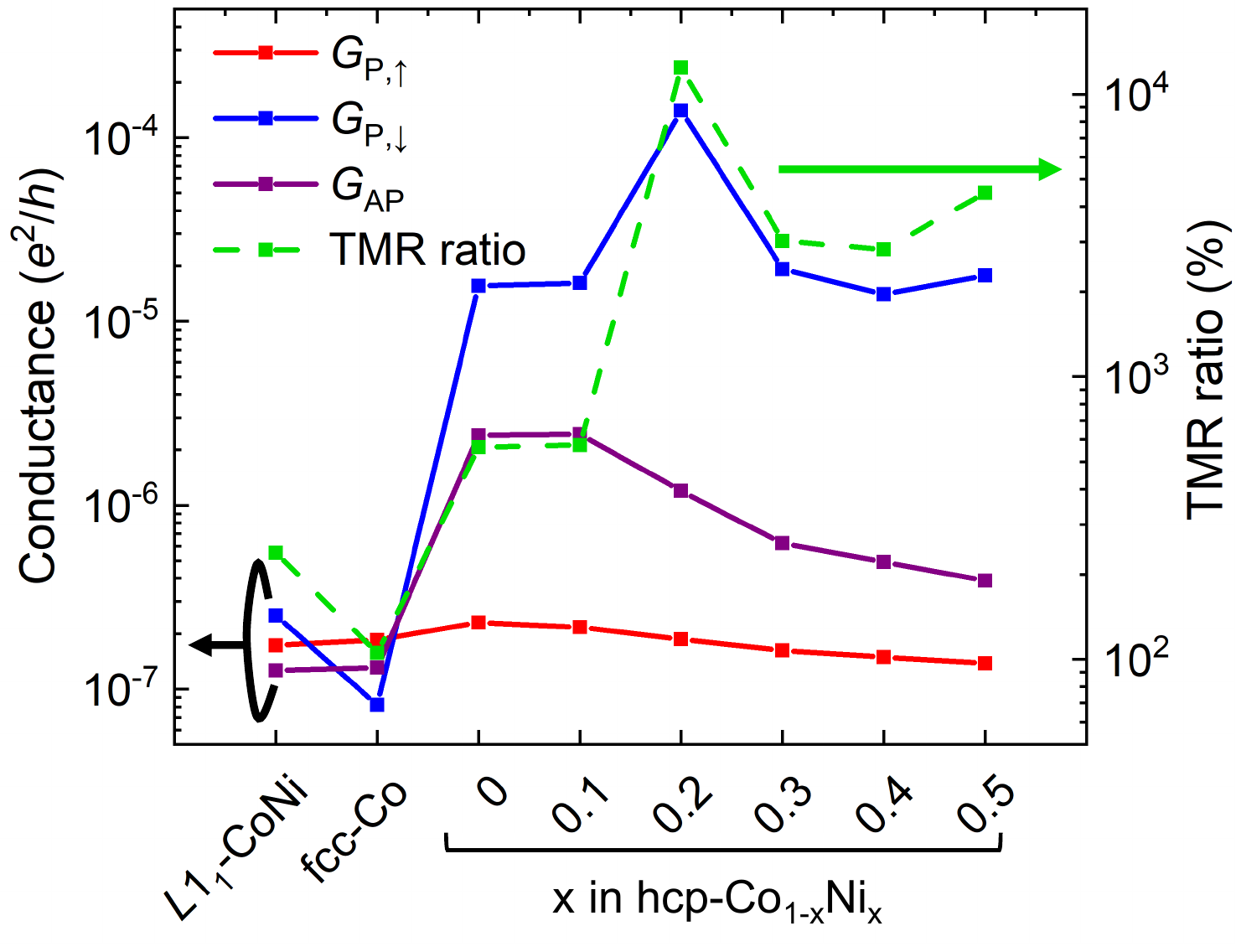}
\caption{\label{fig5} 
Structure- and composition-dependent conductances and TMR ratio (green) for \textit{X}$/$\textit{h}-BN$/$\textit{X}, where \textit{X} = $L1_1$-CoNi, fcc-Co, and \ce{hcp-Co_{1-x}Ni_{x}}. Conductances for $G_{\mathrm{P},\uparrow}$, $G_{\mathrm{P},\downarrow}$, and $G_{\mathrm{AP}}$ are shown in red, blue, and purple, respectively.
}
\end{figure}

\Fref{fig5} shows the variation in spin-resolved conductance and TMR ratio across different electrode structures and compositions. As anticipated from the $\mathbf{k}_\parallel$-resolved conductance maps in \Fref{fig3}, the down-spin conductance in the parallel configuration is significantly higher for the hcp-Co electrode compared to fcc-Co and $L1_1$-CoNi, as shown also in \Tref{table1}. Upon substituting Co with Ni in the hcp structure, the up-spin conductance in the parallel configuration decreases slightly, while the antiparallel conductance drops more substantially. The down-spin conductance in the parallel configuration remains nearly unchanged across the doping range, except for a sudden enhancement at $x = 0.2$ in \ce{hcp-Co_{1-x}Ni_x}, which will be discussed later. These trends result in a marked increase in TMR ratio, reaching nearly an order of magnitude higher than that of pure hcp-Co upon 50\% Ni substitution.

\begin{figure}
\centering
\includegraphics[width=\columnwidth]{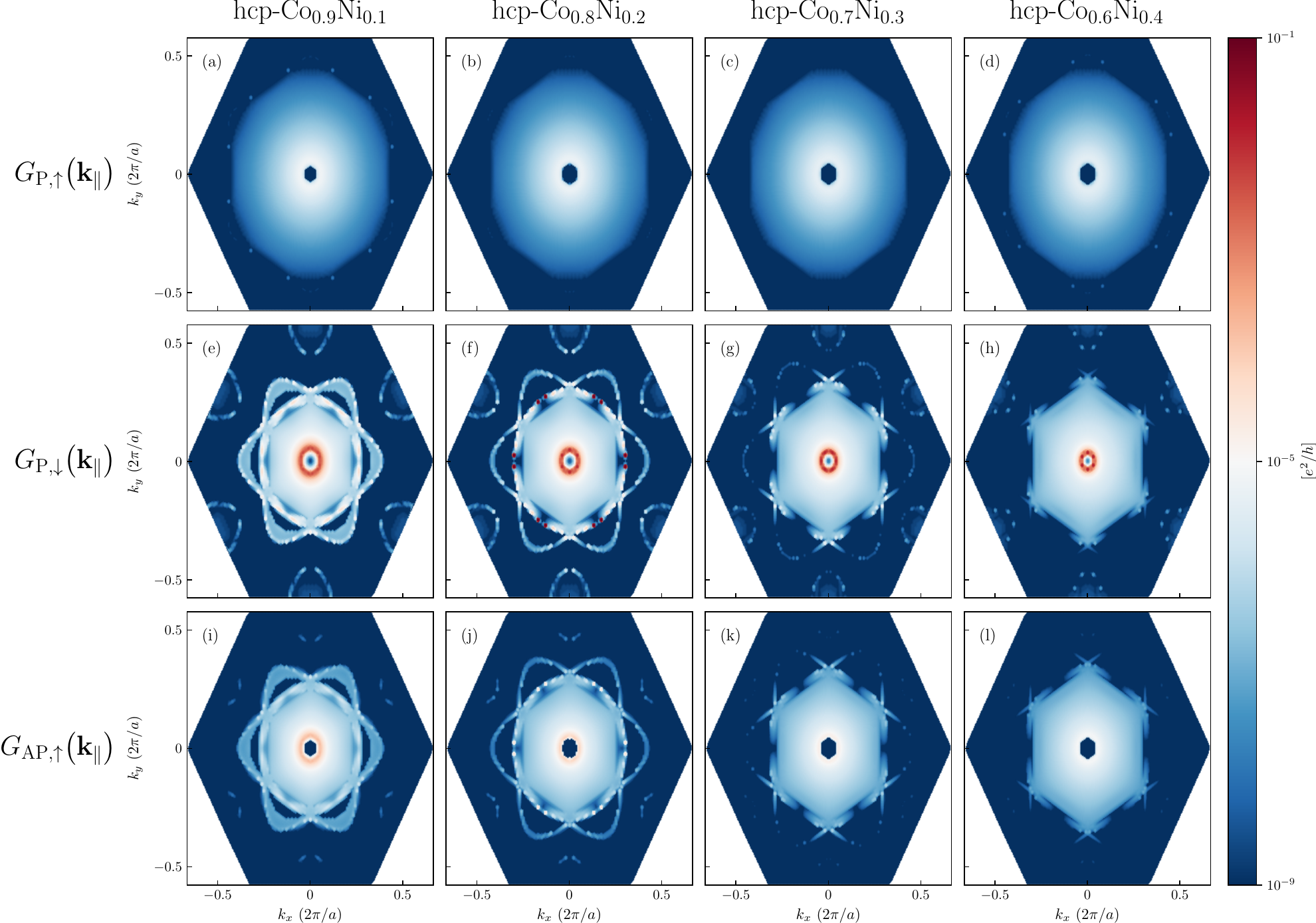}
\caption{\label{fig6} 
$\mathbf{k}_{\parallel}$-resolved conductances at the Fermi level: (a)–(d) up-spin $G_{\mathrm{P},\uparrow}$, (e)–(h) down-spin $G_{\mathrm{P},\downarrow}$ in the parallel configuration, and (i)–(l) up-spin $G_{\mathrm{AP},\uparrow}$ in the antiparallel configuration for hcp-\ce{Co_{1-x}Ni_{x}}$/$\textit{h}-BN$/$hcp-\ce{Co_{1-x}Ni_{x}}(0001) with $x = 0.1$ (a, e, i), $x = 0.2$ (b, f, j), $x = 0.3$ (c, g, k), and $x = 0.4$ (d, h, l).
}
\end{figure}

To understand the microscopic mechanism behind the enhanced TMR ratio upon Ni doping, we examine the $\mathbf{k}_\parallel$-resolved conductance maps for \ce{hcp-Co_{1-x}Ni_{x}} with $0.1 \leq x \leq 0.4$ in \Fref{fig6}(a)–(l). We begin with the up-spin conductance in the parallel configuration [\Fref{fig6}(a)–(d)]. Although the overall profile remains similar across compositions, the zero-conductance region near the $\Gamma$ point gradually expands with increasing Ni content, leading to a slight suppression of up-spin conductance. In contrast, the down-spin conductance in the parallel state [\Fref{fig6}(e)–(h)] exhibits a more pronounced change. The ring-shaped high-conductance feature near the $\Gamma$ point becomes more concentrated toward the center as Ni content increases. Notably, in \ce{hcp-Co_{0.8}Ni_{0.2}} [\Fref{fig6}(f)], an additional hotspot appears midway along the $\Gamma$–K path, which likely accounts for the sudden increase in down-spin conductance observed in \Fref{fig5}.

The trends in the antiparallel configuration [\Fref{fig6}(i)–(l)] can be interpreted as resulting from the mixture of conductances from the up- and down-spin channels in the parallel state. 
In pure hcp-Co, the zero-conductance region of the up-spin channel does not fully overlap with the high-conductance area of the down-spin channel, leaving a finite conductance region near $\Gamma$ [\Fref{fig3}(g)]. 
However, with increasing Ni substitution, the up-spin zero-conductance zone expands [from \Fref{fig6}(a) to 6(d)] and the down-spin high-conductance zone shrinks [from \Fref{fig6}(e) to 6(h)], leading to a substantial reduction in antiparallel conductance. 
\textcolor{black}{Note that the effect of the hotspots along the $\Gamma$–K path is still weakly visible as tiny white spots in \Fref{fig6}(j), but it does not significantly affect the overall reduction of the antiparallel conductance.} 
The reduction directly contributes to the TMR enhancement. 
These results exemplify a form of Brillouin-zone (BZ) spin filtering, as originally proposed by Faleev \textit{et al.}, in which high transmission occurs selectively for one spin channel within a confined region of the BZ~\cite{faleev2015}. 
While the effect is not idealized here, it offers a practical pathway for TMR enhancement in hcp-Co$/$\textit{h}-BN(0001)–based MTJs without altering the chemically inert \textit{h}-BN barrier. 
\textcolor{black}{At the practical level, Ni doping in real materials inevitably introduces defects and breaks the ideal hcp symmetry, which would reduce the surface-state contribution to the TMR ratio. 
This contrasts with our VCA treatment, which assumes a uniform Co–Ni mixing. 
Therefore, the experimentally observed TMR ratio for Ni-doped hcp-Co electrodes may be somewhat lower than the value predicted by our VCA calculations.}

\begin{figure}
\centering
\includegraphics[width=\columnwidth]{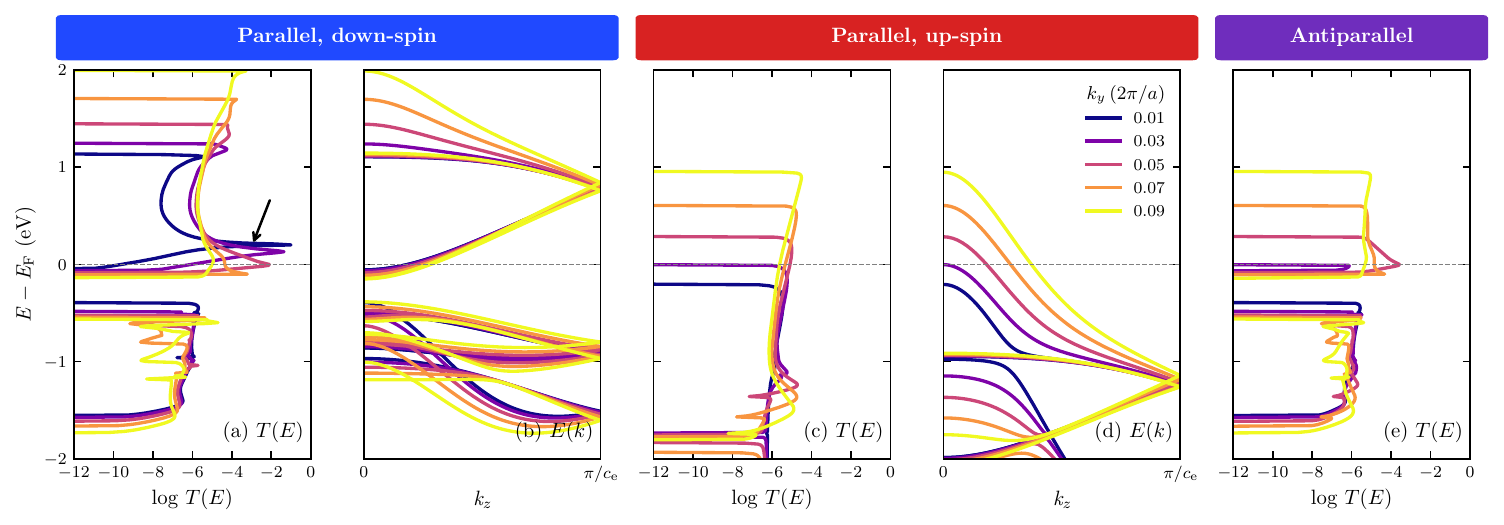}
\caption{\label{fig7} 
Energy-dependent transmittance and band structure along the $k_z$ direction in hcp-Co$/$\textit{h}-BN$/$hcp-Co(0001) MTJs for various $k_y$ values ($\mathbf{k}_{\parallel} = (0, k_{y})$, with $k_y = 0.01–0.09$): (a) total transmittance for down-spin in the parallel configuration; (b) corresponding down-spin band structure; (c), (d) same as (a), (b) for up-spin; (e) same as (a) for the antiparallel configuration. \textcolor{black}{The unit of $k_z$, $\pi / c_\mathrm{e}$, is defined using the lattice constant of the ferromagnetic electrode along the $c$ axis, as illustrated in Fig. 1.}
}
\end{figure}

In previous subsection, the $\mathbf{k}_\parallel$-resolved conductance at the Fermi level was explained qualitatively by differences in the electrode band structure along both high-symmetry ($\Gamma$–A) and lower-symmetry ($\Gamma^*$–A$^*$, with $\mathbf{k}_\parallel = (0, 0.05)$) directions. To gain deeper insight, in this subsection we focus on what happens at $\mathbf{k}_\parallel = (0, k_y)$. \textcolor{black}{Note that the choice of varying $k_y$ is made purely for clarity, as an equivalent trend is expected if $k_y$ were fixed and $k_x$ varied instead.}. \Fref{fig7} presents the energy-dependent \textcolor{black}{spin-resolved} transmittance \textcolor{black}{$T^{\sigma}(\mathbf{k}_\parallel,E)$} and corresponding band structures along the $\Gamma^*$–A$^*$ path, where $\mathbf{k}_\parallel = (0, k_y)$ and $k_y$ is varied from 0.01 to 0.09 for the hcp-Co electrode. As shown in \Fref{fig7}(a)–(d), the energy ranges where finite transmittance appears in the down-spin and up-spin channels under parallel magnetization correlate well with the presence of propagating bands exhibiting $\Delta^*_1$ symmetry. As expected, the antiparallel transmittance [\Fref{fig7}(e)] primarily arises from the energy overlap between the spin-resolved conductances of the parallel configuration. Interestingly, a prominent transmittance peak emerges in the down-spin parallel configuration near the Fermi level [indicated by the black arrow in \Fref{fig7}(a)], despite no clear corresponding feature in the band structure dispersion [\Fref{fig7}(b)]. \textcolor{black}{Here, the transmittance peak is defined as a local maximum of $T^{\sigma}(\mathbf{k}_\parallel,E)$ with respect to $E$, for fixed $\sigma$ and $\mathbf{k}_\parallel$.} This peak becomes sharper and more pronounced as $k_y$ approaches zero, suggesting a possible resonance or concentrated tunneling channel. At larger $k_y$, such as 0.05, the peak broadens and aligns with the enhanced conductance observed at the Fermi level in \Fref{fig3}(d). We will revisit the possible origin of this peak and its dependence on the interfacial distance in the following subsection.

\begin{figure}
\centering
\includegraphics[width=\columnwidth]{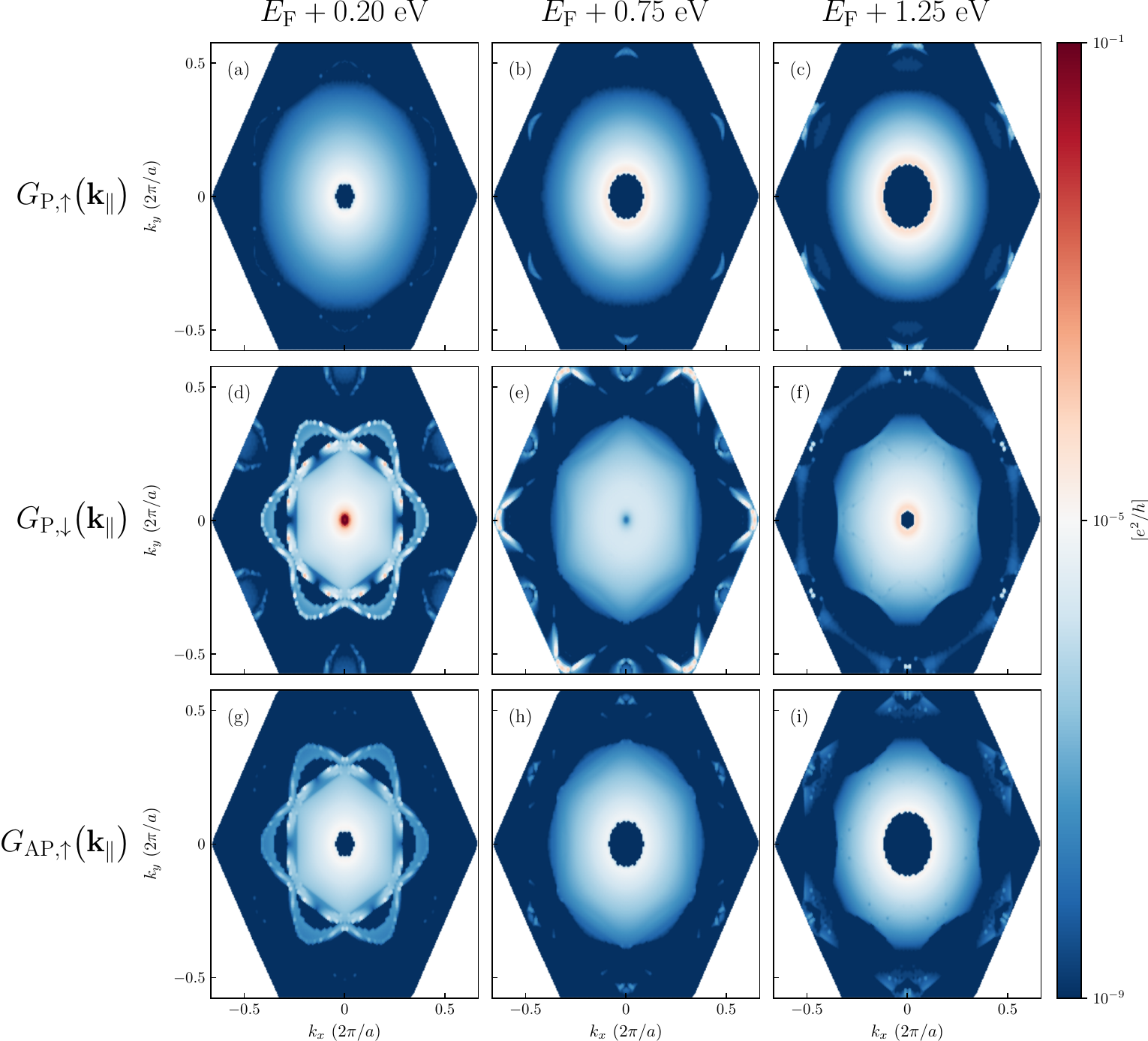}
\caption{\label{fig8} 
$\mathbf{k}_\parallel$-resolved conductances of hcp-Co$/$\textit{h}-BN$/$hcp-Co(0001) MTJs at elevated energies above the Fermi level: (a)–(c) up-spin $G_{\mathrm{P},\uparrow}$, (d)–(f) down-spin $G_{\mathrm{P},\downarrow}$ in the parallel configuration, and (g)–(i) up-spin $G_{\mathrm{AP},\uparrow}$ in the antiparallel configuration at $E_\mathrm{F}+0.2$ eV (a, d, g), $E_\mathrm{F}+0.75$ eV (b, e, h), and $E_\mathrm{F}+1.25$ eV (c, f, i).
}
\end{figure}

A sharp transmittance peak appears at $k_y = 0.01$ around 0.2 eV above the Fermi level, prompting us to examine the $\mathbf{k}_\parallel$-resolved conductance at this energy and higher levels, as shown in \Fref{fig8}. The overall behavior for up-spin conductance in both the parallel and antiparallel configurations remains consistent with the Ni-doping trends [\Fref{fig8}(a)–(c) and (g)–(i)], where the zero-conductance region surrounding the $\Gamma$ point gradually expands. In the down-spin channel [\Fref{fig8}(d)–(f)], the high-conductance region becomes most tightly concentrated around the $\Gamma$ point (though not exactly at $\Gamma$ due to symmetry constraints) at 0.2 eV. At higher energies, such as 0.75 eV, the conductance around $\Gamma$ is more moderate, consistent with the dip observed in the energy-dependent transmittance plots of \Fref{fig7}(a), while additional channels emerge near the Brillouin zone edges. At 1.25 eV, strong conductance re-emerges around the $\Gamma$ point, consistent with the upper transmittance and band edges observed for $k_y \geq 0.05$ in \Fref{fig7}(a). These results support the idea that electron doping remains effective to improve TMR ratio as long as the high-conductance region can still be concentrated near the $\Gamma$ point—an aspect we elaborate further in the following discussion. 

\begin{figure}
\centering
\includegraphics[width=\columnwidth]{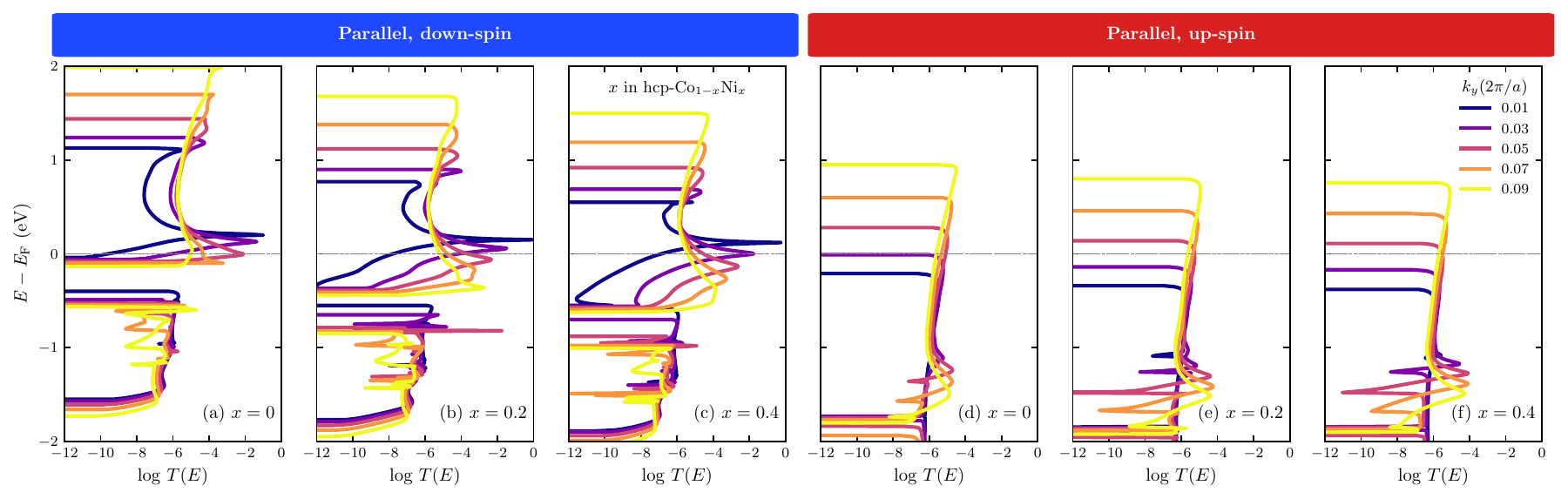}
\caption{\label{fig9} 
Same as \Fref{fig7}, showing energy-dependent transmittance for hcp-\ce{Co_{1-x}Ni_{x}}$/$\textit{h}-BN$/$hcp-\ce{Co_{1-x}Ni_{x}}(0001) MTJs: (a)–(c) down-spin and (d)–(f) up-spin channels of parallel configuration for $x = 0$ (a, d), $x = 0.2$ (b, e), and $x = 0.4$ (c, f).
}
\end{figure}

Nickel has one more valence electron than cobalt, so increasing the Ni concentration effectively raises the Fermi level of the electrode [\Fref{fig4}(a), (d), (e)]. This corresponds to a downward energy shift of the minority-spin $\Delta^*_1$ band that crosses the Fermi level, as seen in the band structures for higher Ni compositions [\Fref{fig4}(f), (i), (j)]. A similar trend is reflected in the transmittance edge positions, as shown in \Fref{fig9}(a)–(f). Interestingly, the strong transmittance peak observed near $k_y = 0.01$ for the down-spin channel in the parallel configuration remains at nearly the same energy, regardless of Ni content [\Fref{fig9}(a)–(c)]. For larger $k_y$ values, the peak becomes broader. Because the minimum of the minority-spin $\Delta^*_1$ band shifts to lower energies with increasing Ni concentration, the transmittance peak begins to cross the Fermi level at smaller $k_y$ values—that is, closer to the $\Gamma$ point. This explains the observed concentration of the down-spin conductance profile toward $\Gamma$ in the $\mathbf{k}_\parallel$-resolved maps as $x$ increases [\Fref{fig6}(e)–(h)]. In contrast, the up-spin channel behavior is more straightforward. As the $\Delta_1$-like band and corresponding transmittance edge shift downward in energy with increasing $x$, the critical $k_y$ value where transmission becomes zero also increases. This leads to an expansion of the zero-conductance region around $\Gamma$ in the up-spin $\mathbf{k}_\parallel$-resolved maps [\Fref{fig6}(a)–(d)]. This overall trend is consistent with the energy-dependent transmittance calculations as a function of $k_y$, where the antiparallel channel follows the overlapping pattern of both spin channels in the parallel configuration (not shown). These findings suggest that the enhancement of the TMR ratio with increasing Ni content cannot be fully explained by changes in the bulk band structure alone. Instead, it involves a subtle alignment between the transmittance peak and the shifting Fermi level, which alters the $\mathbf{k}_\parallel$-resolved transmission landscape.

\subsection{Interfacial distance dependence}

\begin{figure}
\centering
\includegraphics[width=\columnwidth]{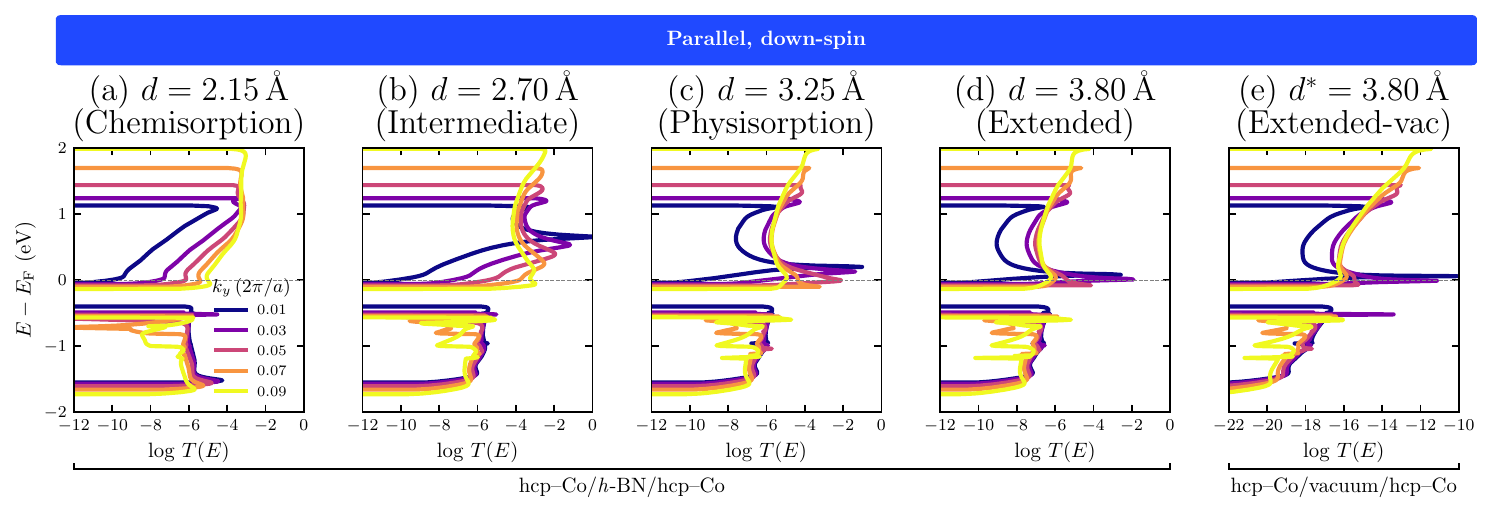}
\caption{\label{fig10} 
Same as \Fref{fig7}, showing down-spin transmittance of hcp-Co$/$\textit{h}-BN$/$hcp-Co(0001) for varying interfacial distances: $d =$ 2.15 Å (a), 2.70 Å (b), 3.25 Å (c), and 3.80 Å (d). (e) Same as (d) but with the 5-ML \textit{h}-BN barrier fully removed, forming a hcp-Co$/$vacuum$/$hcp-Co(0001) junction.
}
\end{figure}

In the previous subsection, we highlighted a prominent parallel down-spin transmittance peak near the Fermi level [\Fref{fig9}(a)–(c)], which plays a key role in the high TMR ratio observed with hcp-Co-based electrodes. In this section, we investigate how this peak is affected by the interfacial distance between the electrode and the \textit{h}-BN barrier. As shown in \Fref{fig10}(a)–(c), reducing the interfacial distance shifts the peak to higher energies and eventually causes it to vanish once the system enters the chemisorption regime. The peak, however, remains present even when the interfacial distance exceeds the physisorption distance [\Fref{fig10}(d)]. A similar feature is also observed in the absence of the \textit{h}-BN barrier, as demonstrated in the hcp-Co$/$vacuum$/$hcp-Co(0001) junction [\Fref{fig10}(e)], although the overall transmittance in this case is significantly lower than in the \textit{h}-BN system.

We considered the possibility that this peak arises from interfacial resonance, where hybridization between adjacent atoms enhances transmission~\cite{masuda2020,masuda2021}. However, this scenario does not align with our findings. First, the peak becomes more pronounced as the interfacial distance increases, which corresponds to weaker hybridization. Second, at shorter distances, where hybridization is expected to be stronger, the peak shifts to higher energy and eventually disappears. Previous studies have also shown that interfacial-resonance-driven tunneling is strongly suppressed when the barrier is replaced with vacuum~\cite{masuda2020}. In contrast, our results show that the sharp peak remains even in the vacuum case, suggesting that a different mechanism is responsible. We further examined the robustness of the peak under various structural modifications. Changing the interface termination from B-top to N-top caused no significant difference in its position or intensity. Increasing the \textit{h}-BN thickness showed that the peak shape converges beyond three monolayers, with the most visible changes occurring between one and three layers. The peak is also preserved when the \textit{h}-BN barrier is fully replaced by a vacuum gap. These results indicate that the peak does not originate from the barrier material, but rather from intrinsic properties of the hcp-Co electrode, likely surface states that couple efficiently to evanescent modes near the Fermi level.

\begin{figure}
\centering
\includegraphics[width=\columnwidth]{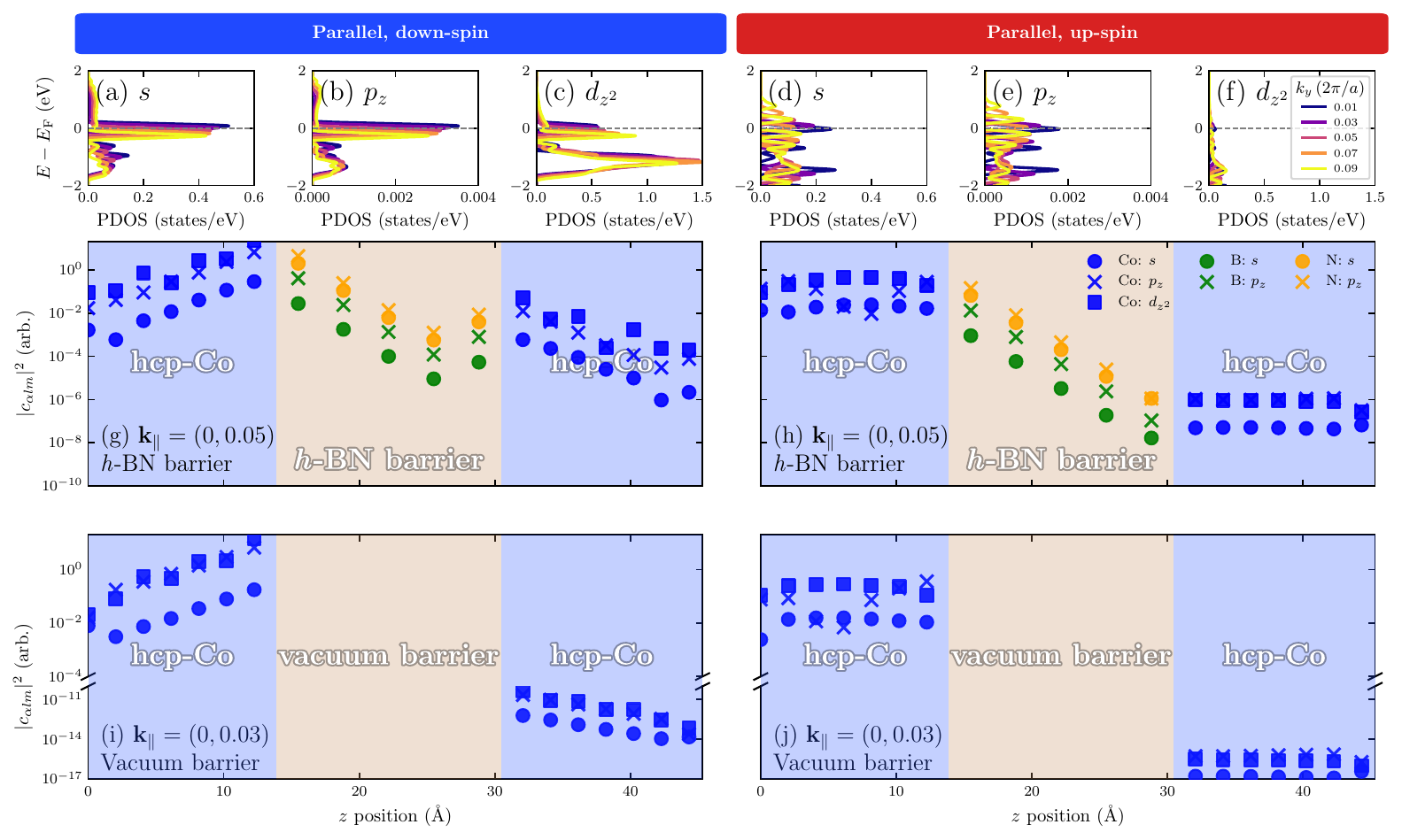}
\caption{\label{fig11} 
Density of states (DOS) for interfacial Co atoms at the $(0, k_y, 0)$ k-point for (a)–(c) down-spin and (d)–(f) up-spin components: (a, d) $s$ orbital, (b, e) $p_z$ orbital, and (c, f) $d_{z^2}$ orbital. Panels (g) and (h) show the projection of the scattering wave function onto local atomic orbitals ($|c_{\alpha lm}|^2$) for the $s$, $p_z$, and $d_{z^2}$ orbitals, plotted as a function of $z$ position across the hcp-Co$/$\textit{h}-BN$/$hcp-Co supercell with a physisorption-type interfacial distance, under the parallel magnetization configuration at $\textbf{k}_\parallel = (0, 0.05)$: (g) down-spin and (h) up-spin channels. Panels (i) and (j) show the corresponding results for the hcp-Co$/$vacuum$/$hcp-Co system at $\textbf{k}_\parallel = (0, 0.03)$.
}
\end{figure}

We analyzed the interfacial density of states (DOS) of Co atoms for the $\Delta^*_1$ symmetry states across various $k_y$ values in both spin channels, as shown in \Fref{fig11}(a)–(f). In the down-spin channel, localized surface states are clearly observed near the Fermi level, whereas such states are absent in the up-spin channel of hcp-Co. Additional evidence is provided by the projection of the scattering wave function onto local atomic orbitals, evaluated at the Fermi level in terms of $\vert c_{\alpha lm}\vert^2$, where $\alpha$, $l$, and $m$ denote the atomic site and orbital quantum numbers, respectively, as defined in Ref.~\cite{choi1999}. For MTJs with an \textit{h}-BN barrier, the highest  transmittance for the parallel down-spin occurs at $\textbf{k}_\parallel = (0, 0.05)$. The corresponding values of $\vert c_{\alpha lm}\vert^2$ are shown in \Fref{fig11}(g) and (h) for the parallel down-spin and up-spin configurations, respectively. In the down-spin case, $\vert c_{\alpha lm}\vert^2$ increases from the electrode side into the left-hand side of the interface region, indicating a strong contribution from surface-localized states. In contrast, the up-spin channel shows nearly constant $\vert c_{\alpha lm}\vert^2$ across the same region, suggesting a lack of surface state contribution. A similar trend is observed in the vacuum-junction configuration, where the highest transmittance for the parallel down-spin occurs at $\textbf{k}_\parallel = (0, 0.03)$ [\Fref{fig11}(i)–(j)]. This confirms that such down-spin surface states can exist even in the absence of a barrier and may significantly contribute to the high TMR ratio when the interfacial separation is sufficiently large.

This mechanism resembles surface-state-assisted tunneling previously reported in MgO-based MTJs, where down-spin surface states play a critical role in transmission~\cite{rungger2007,rungger2009,waldron2006,wunnicke2002}. We also find that the TMR ratio in the hcp-Co$/$vacuum$/$hcp-Co(0001) junction is higher than in the corresponding \textit{h}-BN–based junction, further supporting the surface-state origin of the transmittance peak. Although the barrier material may influence the energetic position and amplitude of these states, it does not appear to be the main source of the peak. These results indicate that hcp-Co-based MTJs can maintain a high TMR ratio even when combined with other two-dimensional insulators and semiconductors, as long as the interfacial distance is sufficiently large to preserve the surface-state contribution. This insight may guide the design of efficient spintronic devices using hcp-Co alloys ferromagnet and 2D insulator or semiconductor materials that allow for weak interface interaction.

\section{Conclusion}
In this study, we elucidate the origin of the high tunnel magnetoresistance (TMR) ratio predicted for hcp-Co$/$\textit{h}-BN$/$hcp-Co(0001) magnetic tunnel junctions (MTJs) with physisorption-type interfaces. We identify a $\Delta_1$-like band along an off-symmetry $\Gamma^*$–A$^*$ path that plays a central role in enhancing the down-spin conductance—an effect not observed in other electrode structures such as fcc-Co and $L1_1$-CoNi. A transmittance peak emerges in the down-spin channel near the $\Gamma$ point just above the Fermi level. Upon Ni doping, this peak broadens and shifts toward the Fermi level, resulting in a more concentrated high-conductance region around $\Gamma$. Simultaneously, the zero-conductance regions in the up-spin and antiparallel configurations expand. These changes collectively enhance the TMR ratio through a mechanism analogous to Brillouin zone spin filtering. Further analysis indicates that the transmittance peak originates from surface states of hcp-Co, whose energy position is sensitive to the interaction with the \textit{h}-BN layer. As the interfacial distance decreases toward the chemisorption regime, the peak disappears, suggesting that high TMR ratios are attainable only under physisorption conditions. This finding opens a path to exploit surface-state-assisted tunneling in hcp-Co–based MTJs using alternative two-dimensional insulating or semiconducting barriers, provided the interfacial distance is sufficiently large to preserve the surface state contribution.

\section{Data availability statement}
All data that support the findings of this study are included
within the article (and any supplementary files).

\ack{The authors are grateful to T. Tadano, E. Xiao, H. Sukegawa of NIMS, and H. Ago of Kyushu University for their valuable discussions on this paper. This research was partially supported by Grants-in-Aid for Scientific Research (Grant No. JP22H04966 and JP23K03933) from the Japan Society for the Promotion of Science. The calculations were performed on the Numerical Materials Simulator at NIMS.}

\section*{References}

\end{document}